\documentclass{easychair}

\usepackage{doc}
\usepackage{makeidx}
\usepackage[leqno,fleqn,intlimits]{amsmath}
\usepackage{graphicx}
\usepackage{color}
\usepackage{float}
\usepackage{url}
\usepackage[utf8]{inputenc}
\usepackage[T1]{fontenc}

\definecolor{red_ex}{rgb}{0.7,0,0}
\definecolor{blue_ex}{rgb}{0,0,0.7}

\begin{document}


\title{ {\sc Artex} {\sl is AnotheR TEXt summarizer}}


\titlerunning{ {\sc Artex} {\sl is AnotheR TEXt summarizer}}

%
\author{
Juan-Manuel {\sc Torres-Moreno}\inst{1}$^,$\inst{2}
}

\institute{
  Laboratoire Informatique d'Avignon,\\
  BP 91228 84911, 
  Avignon, Cedex 09, France\\
  \email{juan-manuel.torres@univ-avignon.fr}
\and
   \'Ecole Polytechnique de Montr\'eal,\\ CP. 6128 succursale Centre-ville,
   Montréal, Québec, Canada\\
 }

\authorrunning{\sc Torres-Moreno}

\clearpage

\maketitle

\begin{abstract}
This paper describes {\sc Artex}, another algorithm for Automatic Text Summarization.
In order to rank sentences, a simple inner product is calculated between each sentence, a document vector (text topic) 
and a lexical vector (vocabulary used by a sentence).
Summaries are then generated by assembling the highest ranked sentences.
No ruled-based linguistic post-processing is necessary in order to obtain summaries.
Tests over several datasets (coming from Document Understanding Conferences (DUC), Text Analysis Conference (TAC), evaluation campaigns, etc.) in French, English and Spanish  
have shown that {\sc Artex} summarizer achieves interesting results.
\end{abstract}


\setcounter{tocdepth}{2}

\pagestyle{empty}

\section{Introduction}
\label{sec:intro}

Automatic Text Summarization (ATS) is the process to automatically generate a compressed version of a source document \cite{torres:11}.
Query-oriented summaries focus on a user's request, and extract the information related to the specified topic given explicitly in the form of a query \cite{daume:06}.
Generic mono-document summarization tries to cover as much as possible the information content.
Multi-document summarization is a oriented task to create a summary from a heterogeneous set of documents on a focused topic.
Over the past years, extensive experiments on query-oriented multi-document summarization have been carried out.
Extractive Summarization produces summaries choosing a subset of representative sentences from original documents.
Sentences are ordered and then assembled according to their relevance to generate the final summary \cite{mani:mayburi:99}.

This article introduces a new method of summarization based in sentences extraction on Vector Space Model (VSM).
We score each sentence by calculating their inner product with a pseudo-sentence vector and a pseudo-word vector.
Results show that {\sc Artex} not only preserves the content of the summaries generated using this new representation, but often, 
surprisingly the performance can be improved.
{\sc Artex} could be an interesting and simple algorithm using the extractive summarization paradigm.
Our tests on trilingual corpora (English, Spanish and French) evaluated by the {\sc Fresa} algorithm (without human references) confirm the good performance of {\sc Artex}.

In this paper, related work is given in Section \ref{sec:relatedwork}.
Section \ref{sec:artex} presents the new algorithm of Automatic Text Summarization.
Experiments are presented in Section \ref{sec:experiments}, followed by Results in Section \ref{sec:results} and Conclusions in Section \ref{sec:conclusion}.

\section{Related works}
\label{sec:relatedwork}

Research in Automatic Text Summarization was introduced by H.P. Luhn in 1958 \cite{luhn:58}.
In the stra\-tegy proposed by Luhn, the sentences are scored for their component word values as determined by tf*idf-like weights.
Scored sentences are then ranked and selected from the top until some summary length threshold is reached.
Finally, the summary is generated by assembling the selected sentences in the original source order.
Although fairly simple, this extractive methodology is still used in current approaches.
Later on, \cite{edmundson:69} extended this work by adding simple heuristic features such as the position of sentences in the text or some key phrases indicate 
the importance of the sentences.
As the range of possible features for source characterization widened, choosing appropriate features, feature weights and feature combinations have become a central issue.

A natural way to tackle this problem is to consider sentence extraction as a classification task.
To this end, several machine learning approaches that uses document-summary pairs have been proposed \cite{kupiec:95,teufel:moens:97},
An hybrid method mixing statistical and linguistics algorithms is presented in \cite{iria:micai:07}.
\cite{mani:mayburi:99} and \cite{torres:11} propose a good state-of-art of Automatic Text Summarization tasks and algorithms.

\subsection{Document Pre-processing}
\label{sec:pre-processing}

The first step to represent documents in a suitable space is the pre-processing.
As we use extractive summarization, documents have to be chunked into cohesive textual segments that will be assembled to produce the summary.
Pre-processing is very important because the selection of segments is based on words or bigrams of words.
The choice was made to split documents into full sentences, in this way obtaining textual segments that are likely to be grammatically corrects.
Afterwards, sentences pass through several basic normalization steps in order to reduce computational complexity.

The process is composed by the following steps: 

\begin{enumerate}
\item \textbf{Sentence splitting}. Simple rule-based method is used for sentence splitting. 
	Documents are chunked at the period, exclamation and question mark. 
\item \textbf{Sentence filtering}. Words lowercased and cleared up from sloppy punctuation. 
	Words with less than 2 occurrences ($f<2$) are eliminated ({\sl Hapax legomenon} presents once in a document).
	Words that do not carry meaning such as functional or very common words are removed.
	Small stop-lists (depending of language) are used in this step.
\item \textbf{Word normalization}. Remaining words are replaced by their canonical form using lemmatization, stemming, ultra-stemming or none of them (raw text). 
Four methods of normalization were applied after filtering: 
\begin{itemize}
	\item Lemmatization by simples dictionaries of morphological families. 
	These dictionaries have 1.32M, 208K and 316K  words-entries in Spanish, English and French, respectively.  
	\item Porter's Stemming, available at Snowball (web site \url{http://snowball.tartarus.org/texts/stemmersoverview.html}) for English, Spanish, French among other languages.
  	\item Ultra-stemming.  
	This normalization seems be very efficient and it produces a compact matrix representation \cite{torres:arxiv:12}. 
	Ultra-stemming consider only the $n$ first letters of each word.
	For example, in the case of ultra-stemming (first letter, {\sc Fix$_1$}), inflected verbs like ``sing'', ``song'', ``sings'', ``singing''... or proper names ``smith'', 
	``snowboard'', ``sex'',... are replaced by the letter ``{\bf s}''. 
\end{itemize}

\item \textbf{Text Vectorization}. Documents are vectorized in a matrix $S_{[P \times N]}$ of $P$ sentences and $N$ columns.
	Each element $s_{i,j}$ represents the occurrences 
	of an object $j$ (a letter in the case of ultra-stemming, a word in the case of lemmatization or a stem for stemming), $j=1,2,...,N$ in the sentence $i$, $i=1,2,...,P$. 
\end{enumerate}

\section{{\sl AnotheR TEXt summarizer} ({\sc Artex})}
\label{sec:artex}

\begin{figure}[H]
	\centering
	\includegraphics[width=0.65\textwidth]{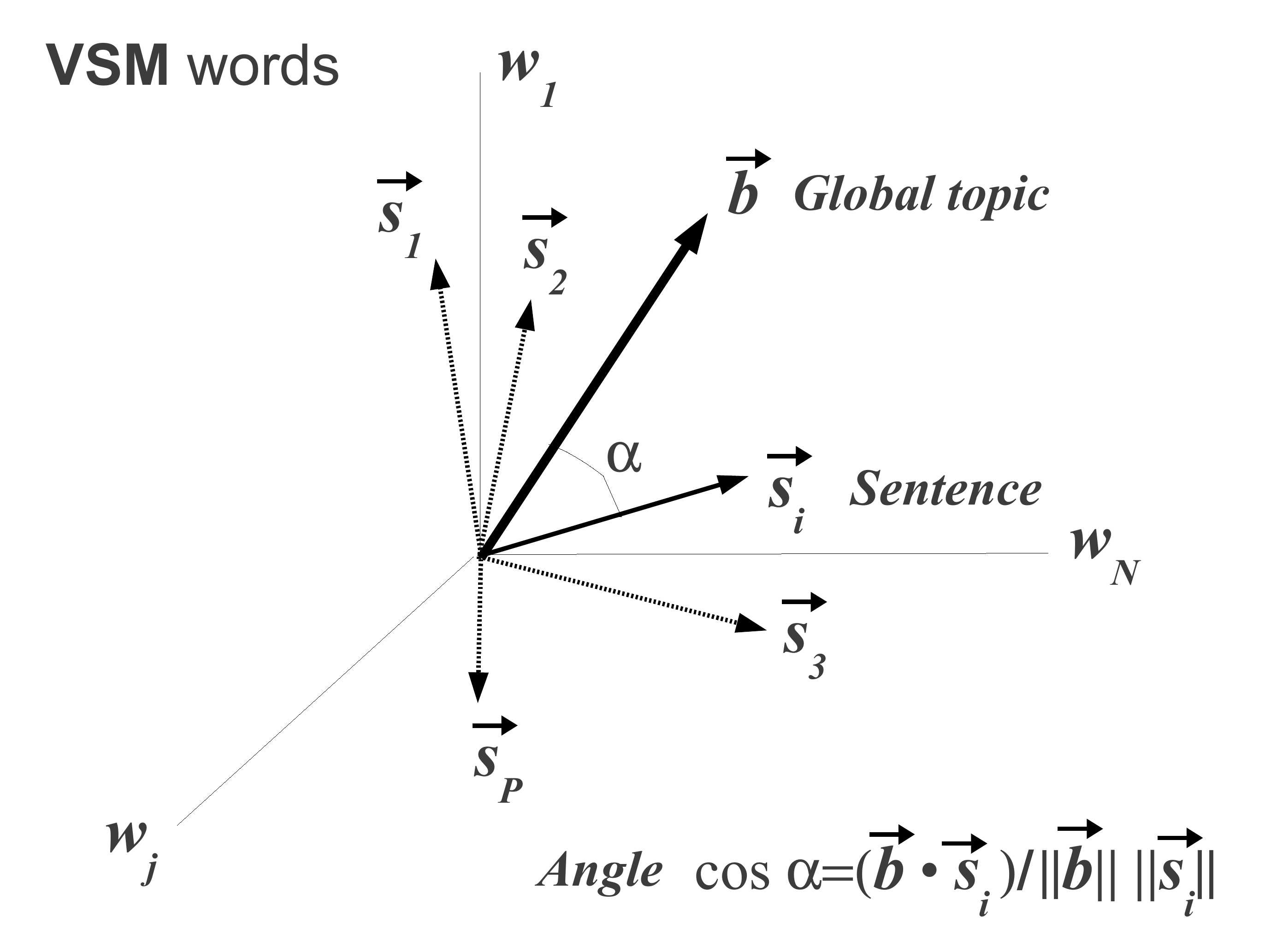}
	\caption{The ``global topic'' in a Vector Space Model of $N$ words.}
	\label{fig:svm}
\end{figure}

	{\sc Artex}\footnote{In French, {\sc Artex} {\sl est un Autre Résumeur TEXtuel}.} is a simple extractive algorithm for Automatic Text Summarization. 
	The main idea is the next one: 
	First, we represent the text in a suitable space model (VSM). 
	Then, we construct an average document vector that represents the average (the ``global topic'') of all sentences vectors. 
	At the same time, we obtain the ``lexical weight'' for each sentence, i.e. the number of words in the sentence. 
	After that, it is calculated the angle between the average document and each sentence; 
	narrow angles indicate that the sentences near of the ``global topic'' should be important and therefore extracted.
	See on the figure \ref{fig:svm} the VSM of words: $P$ vector sentences and the average ``global topic'' are represented in a $N$ dimensional space of words.

\begin{figure}[H]
	\centering
	\includegraphics[width=0.65\textwidth]{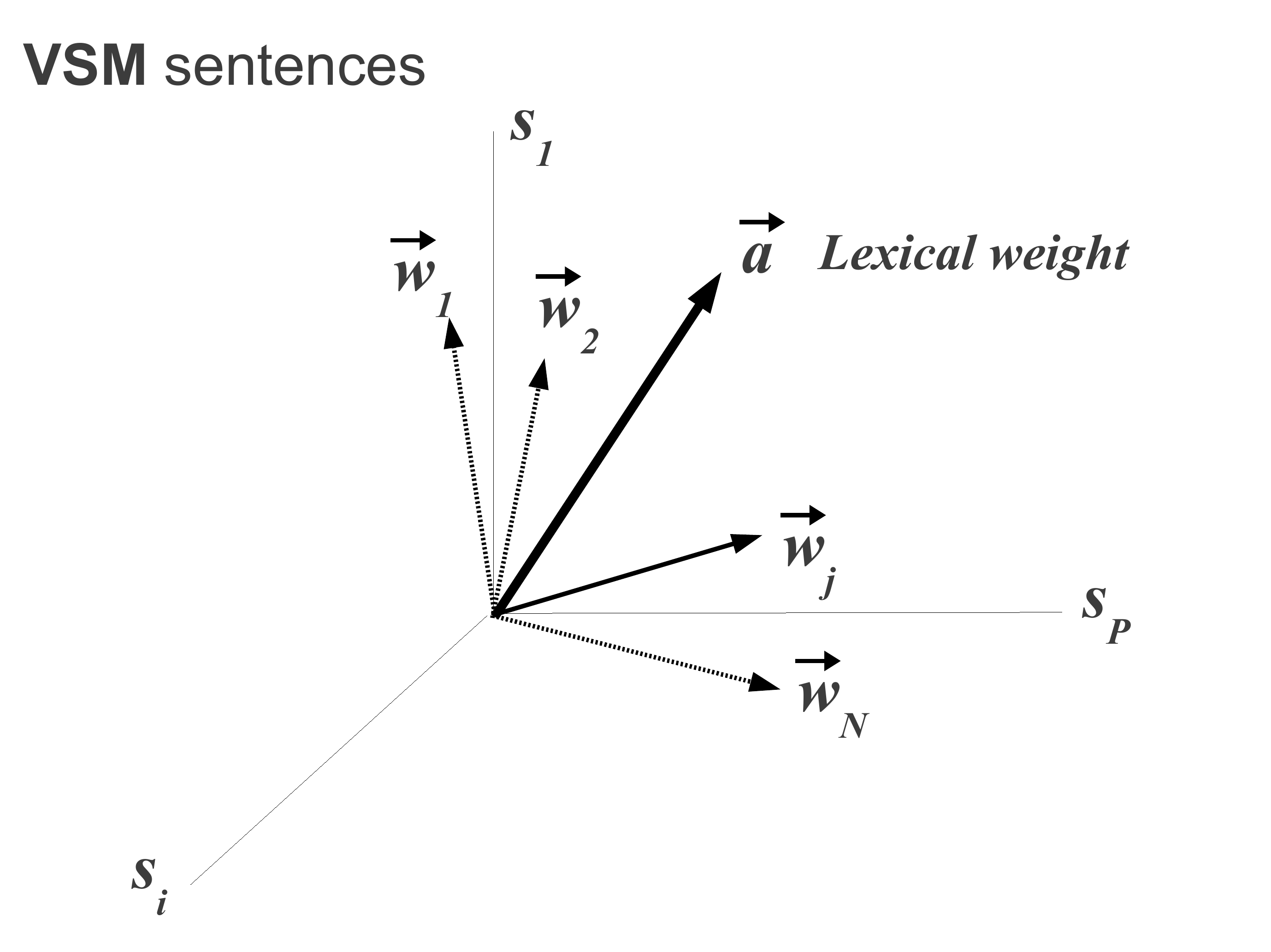}
	\caption{The ``lexical weight'' in a Vector Space Model of $P$ sentences.}
	\label{fig:svm-s}
\end{figure}

	Next, a score for each sentence is calculated using their proximity with the ``global topic'' and their ``lexical weight''.
	In the figure \ref{fig:svm-s}, the ``lexical weight'' is represented in a VSM of $P$ sentences. 

	Finally, the summary is generated concatenating the sentences with the highest scores following their order in the original document.

\subsection{Algorithm}

	Formally, {\sc Artex} algorithm computes the score of each sentence by calculating the inner 
	product between a sentence vector, an {\sl average pseudo-sentence vector} (the ``global topic'') and an {\sl average pseudo-word vector} (the ``lexical weight'').

	Once a pre-processing (word normalization and filtering of stop words) is completed, it is created a matrix $S_{[P \times N]}$, using the Vector Space Model, that contains $N$ words (or letters) and $P$ sentences.

	Let $s_i = (s_1,s_2,...,s_N)$ be a vector of the sentence $i$, $i=1,2,...,P$. 
	We defined ${\vec a}$ the {\sl average pseudo-word vector}, as the average number of occurrences of $N$ words used in the sentence $i$:

	\begin{equation}
		a_i =  \frac{1}{N} \sum_j s_{i,j} 
	\end{equation}

	\noindent and ${\vec b}$ the {\sl average pseudo-sentence vector} as the average number of occurrences of each word $j$ used trough the $P$ sentences:

	\begin{equation}
		b_j =  \frac{1}{P} \sum_i s_{i,j} 
	\end{equation}
	
	The score or weight of each sentence $s_i$ is calculated as follows:

	\begin{equation}
		\textrm{score}(s_i) = \left( \vec{s} \times \vec{b} \right) \ \times \vec{a} = \frac{1}{N P} \left( \sum_j s_{i,j} \times b_j \right) \times a_i \, ; \, i=1,2,...,P \, ; j=1,1,...,N
	\label{eq:artex}
	\end{equation}

	The score($\bullet$) computed by equation \ref{eq:artex} must be normalized between the interval [0,1].
	The calculation of $\vec{s} \times \vec{b}$ indicates the proximity between the sentence $\vec s$ and the {\sl average pseudo-sentence} $\vec b$.
	The product ($\vec{s} \times \vec{b}) \times \vec a$ weigh this proximity using the {\sl average pseudo-word} $a_i$. 

	If a sentence $s_i$ is near of $\vec b$ and their corresponding element $a_i$ has a high value, $s_i$ will have, therefore, a high score.
	Moreover, a sentence $i$ far of main topic (i.e. $\vec{s}_i \times \vec{b}$  is near 0) or a less informative sentence $i$ (i.e. $a_i$ are near 0) will have a low score.

	In computational terms, it is not really necessary to divide the scalar product by the constant $\frac{1}{N P}$, because the angle $\alpha = \arccos {\vec b}.{\vec s}/|{\vec b}||{\vec s}|$ between ${\vec b}$ and $\vec{s}$ is the same
	if we use ${\vec b} = {\vec b}' = \sum_i s_{i,j}$. 
	The element $a_i$ is only a scale factor that does not modify $\alpha$.    

	In fact, if the matrix $S_{[P \times N]}$ is approximated to a binary matrix\footnote{This is a reasonable approximation in this context,  because $S_{[P \times N]}$ is a sparsed matrix with many
	term occurrences equal to one or zero.} $S'_{[P \times N]}$, where each element $s'_{i,j} = \{0,1\}$ has a probability of $p=\frac{1}{2}$, 
	we can normalize vectors $\vec a$, $\vec b$ and matrix $S$, as follows:

	\begin{eqnarray}
		|\vec{a}| &=& \sum_i^P \sqrt{{s'_{i,j}}^2} = \sum_i^P \sqrt{ (\{0,1\}^P)^2}  = N \sqrt{P} \\
		|\vec{b}| &=& \sum_j^N \sqrt{{s'_{i,j}}^2} = \sum_j^N \sqrt{ (\{0,1\}^N)^2}  = \sqrt{N} P \\ 
		|\vec{s}_i| &=& \sum_j^N \sqrt{{s'_{i,j}}^2} = \sum_j^N \sqrt{ \{0,1\}^2}  = N 
	\label{eq:normes}
	\end{eqnarray}

	Vectors then will  be represented in hyper-spheres of $N$ or $P$ dimensions, and the normalized score' in this space would be:

	\begin{eqnarray}
		\textrm{score'}(s_i) &=& \left( \frac{\vec{s}}{|\vec{s}|}  \times \frac{\vec{b}}{|\vec{b}|} \right) \ \times \frac{\vec{a}}{|\vec{a}|} = 
		 \frac{1}{N \sqrt{N} P N \sqrt{P} } \left( \sum_j s_{i,j} \times b_j \right) \times a_i \, \nonumber \\
		 &=&\frac{1}{\sqrt{N^5 P^3}  } \left( \sum_j s_{i,j} \times b_j \right) \times a_i \, ; \, i=1,2,...,P \, ; j=1,2,...,N
	\label{eq:artexnorm}
	\end{eqnarray}
	However, the term  $1/\sqrt{N^5 P^3}$ is a constant value (i.e. a simple scale factor), and then the $\textrm{score}(\bullet)$ calculated using the equation 
	\ref{eq:artex}) and the $\textrm{score'}(\bullet)$ using the equation \ref{eq:artexnorm}, are both equivalent.

\section{Experiments}
\label{sec:experiments}

{\sc Artex} algorithm described in the previous section has been implemented and evaluated in corpora in several languages.

We have conducted our experimentation with the following languages, summarization tasks, summarizers and data sets:
	1) Generic multi-document-summarization in English with the corpus DUC'04; 
	2) Generic single-document summarization in Spanish  with the corpus {\sl Medicina Cl\'inica}  and
	3) Generic single document summarization in French  with the corpus {\sc Pistes}.

We have applied the summarization algorithms and finally, the results have been evaluated using {\sc Fresa}
while processing times for each summarizer have been measured and compared.

The following subsections present formally the details of the summarizers, corpora and evaluations studied in different experiments.

\subsection{Other Summarizers}

To compare the performances, two other summarization systems were used in our experiments: {\sc Cortex} and {\sc Enertex}.
To be in the same conditions, these two systems have used exactly the same textual representation based on Vector Space Model, described in Section \ref{sec:pre-processing}.

\begin{itemize}
	\item {\sc Cortex} is a single-document summarization system using several metrics and an optimal decision algorithm \cite{favre:06,torres:09,torres:11,torres-Moreno2001}.
	\item {\sc Enertex} is a summarization system based in Textual Energy concept \cite{fernandez:micai:07}: text is represented as a spin system where 
	spins $\uparrow$ represents words that their occurrences are $f>1$ (spins $\downarrow$ if the word is not present).
\end{itemize}

\subsection{Summarization Corpora Description}
\label{sec:corpora}

To study the impact of our summarizer, we used corpora in three languages: English, Spanish and French.
The corpora are heterogeneous, and different tasks are representatives of Automatic Text Summarization: 
generic multi-document summary and mono-document guided by a subject.

\begin{itemize}
	\item Corpus in English.
	Piloted by NIST in Document Understanding Conference\footnote{\url{http://duc.nist.gov}} (DUC) the Task 2 of DUC'04\footnote{\url{http://www-nlpir.nist.gov/projects/duc/guidelines/2004.html}}, aims to produce a short summary of a cluster of related documents.
	We studied generic multi-document-summarization in English using data from DUC'04.
	This corpus with 300K words (17 780 types) is compound of 50 clusters, 10 documents each. 
	\item Corpus in Spanish. 
	Generic single-document summarization using a corpus from the scientific journal {\sl Medicina Cl\'inica}\footnote{\url{http://www.elsevier.es/revistas/ctl_servlet?_f=7032&revistaid=2}}, 
	which is composed of 50 medical articles in Spanish, each one with its corresponding author abstract.
	This corpus contains 125K words (9 657 types).
	\item Corpus in French.
	We have studied generic single-document summarization using the Canadian French Sociological Articles corpus, generated from the journal 
	{\sl Perspectives interdisciplinaires sur le travail et la sant\'e} ({\sc Pistes})\footnote{\url{http://www.pistes.uqam.ca/}}. It contains  50 sociological articles in French, each one with its corresponding author abstract.
	This corpus contains near 400K words (18 887 types).
\end{itemize}

\subsection{Summaries Content Evaluation}

DUC conferences have introduced the ROUGE content evaluation \cite{lin:2004rpa}, wich measures the overlap of $n$-grams between a candidate summary
and reference summaries written by humans.
However, to write the human summaries necessaries for ROUGE is a very expensive task.

Recently metrics without references have been defined and experimented at DUC and Text Analysis Conferences (TAC)\footnote{\url{www.nist.gov/tac}} workshops. 

\textsc{Fresa} content evaluation \cite{torres:10:taln,torres:10poli} is similar to ROUGE evaluation, but human reference summaries are not necessary.
\textsc{Fresa} calculates the divergence of probabilities between the candidate summary and the document source.
Among these metrics, Kullback-Leibler (KL) and Jensen-Shannon (JS) divergences have been widely used by \cite{louis:nenkova:08,torres:10poli} to evaluate the informativeness of summaries.

In this article, we use {\sc Fresa}, based in KL divergence with Dirichlet smoothing, like in the 2010 and 2011 INEX edition \cite{SanJuan:11}, to evaluate the informative content of summaries 
by comparing their $n$-gram distributions with those from source documents.

{\sc Fresa} only considered absolute log-diff between the terms occurrences of the source and the summary. 
Let $T$ be the set of terms in the source. 
For every $t \in T$, we denote by $C_t^T$ its occurrences in the source and $C_t^S$ its occurrences in the summary. 

The {\sc Fresa} package computed the divergence between the document source and the summaries as follows:

\begin{equation}
	{\mathcal D}(T||S) = \sum_{t \in T} \left| \log \left(\frac{C_t^T}{|T|} + 1\right) - \log\left( \frac{C_t^S}{|S|} + 1\right)\right| 
\end{equation}

To evaluate the information content  (the ``quality'') of the generated summaries, after removing stop-words, several automatic measures were computed:
	\textsc{Fresa$_1$} (Unigrams of single stems),
	\textsc{Fresa$_2$} (Bigrams of pairs of consecutive stems),
	\textsc{Fresa$_{SU4}$} (Bigrams with 2-gaps also made of pairs of consecutive stems) and finally,
	$\langle\textsc{Fresa}\rangle$, i.e. the average of all {\sc Fresa} values.

The {\sc Fresa} values (scores) are normalized between 0 and 1.
High {\sc Fresa} values mean less divergence regarding the source document summary, reflecting a greater amount of information content.
All summaries produced by the systems were evaluated automatically using {\sc Fresa} package.

\section{Results}
\label{sec:results}

In this section we present the results for each corpus with different summarizers and the several normalization strategies used.
Based on these results, firstly, we have verified that ultra-stemming improves the performance of summarizers.
Secondly, we show that {\sc Artex} is a system that has a similar performances --in terms of information content and processing times-- to other state-of-art summarizers.

\subsection{Content evaluation}

\begin{itemize}
\item {\bf English corpus.}
Figure \ref{fig:duc04} shows the performance of the three summarizers using {\sc Fix$_1$}, stemming and lemmatization.
Results show that ultra-stemming improves the score of the three automatic summarizer systems. 
{\sc Artex} and {\sc Cortex} expose a similar performances  in information content.

\begin{figure}[H]
	\centering
	\includegraphics[width=0.75\textwidth]{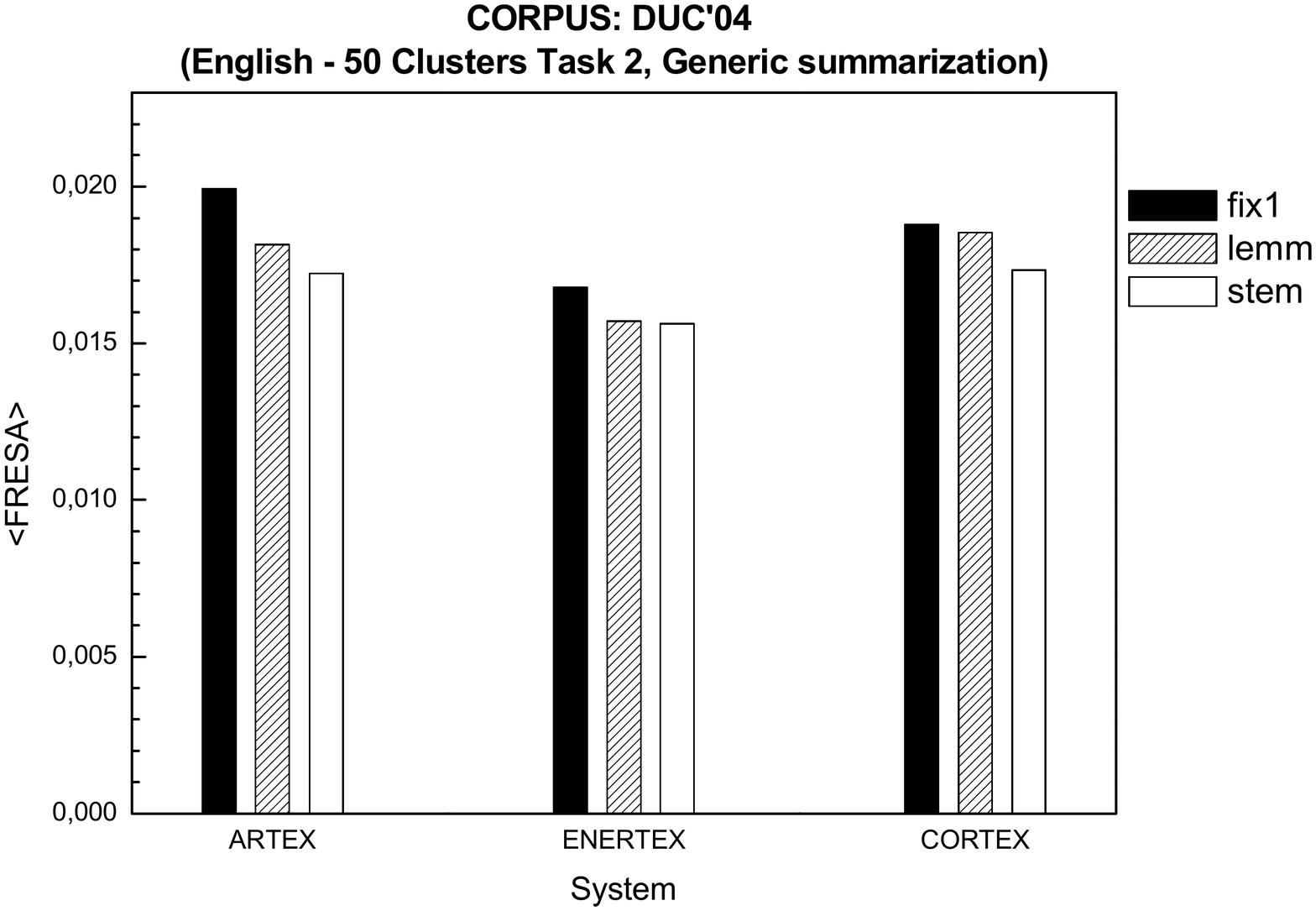}
	\caption{Histogram plot of content evaluation for corpus DUC'04 Task 2, with $\langle$\textsc{Fresa}$\rangle$ measures, for each summarizer and each normalization.}
	\label{fig:duc04}
\end{figure}

\item  {\bf Spanish corpus.}
Spanish is a language with a greater variability than English.
Results in figure \ref{fig:medicina} shown that {\sc Artex} summarizer outperforms {\sc Cortex} and {\sc Enertex} if stemming or lemmatization are used as normalization.

\begin{figure}[H]
	\centering
	\includegraphics[width=0.75\textwidth]{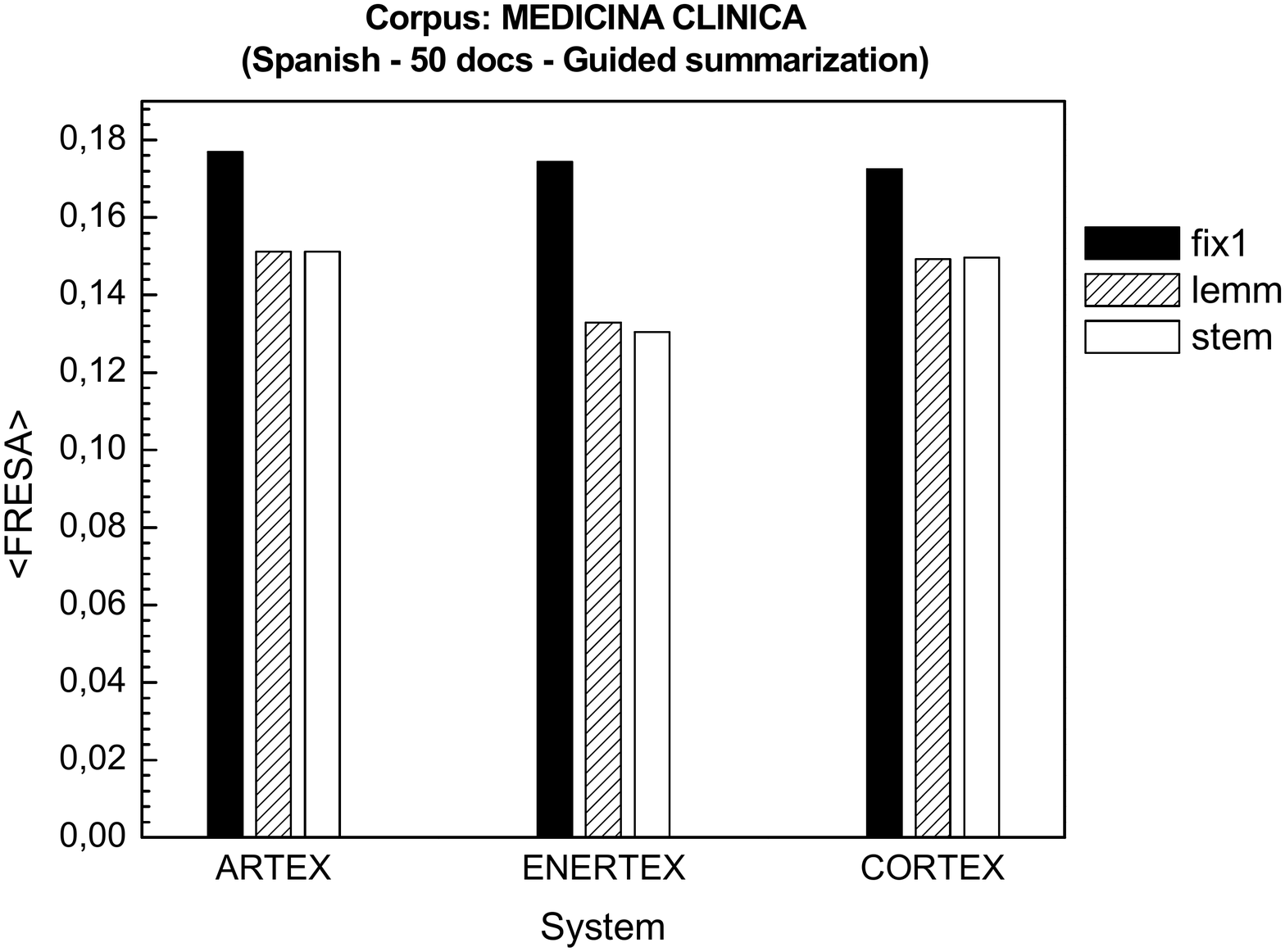}
	\caption{Histogram plot of content evaluation for Spanish corpus {\sl Medicina Cl\'inica} with $\langle$\textsc{Fresa}$\rangle$ scores for each summarizer.}
	\label{fig:medicina}
\end{figure}

\item  {\bf French corpus.}
French is a language with a large variability too. 
Figure \ref{fig:pistes} shows the score $\langle$\textsc{Fresa}$\rangle$ on the French corpus {\sl Pistes}. 
Results show a similar behavior: 
Ultra-stemming improves the score of the three automatic summarization systems used.
In particular, the efficacy of {\sc Artex} is less sensible to word normalization than others summarizers. 
\begin{figure}[H]
	\centering
	\includegraphics[width=0.75\textwidth]{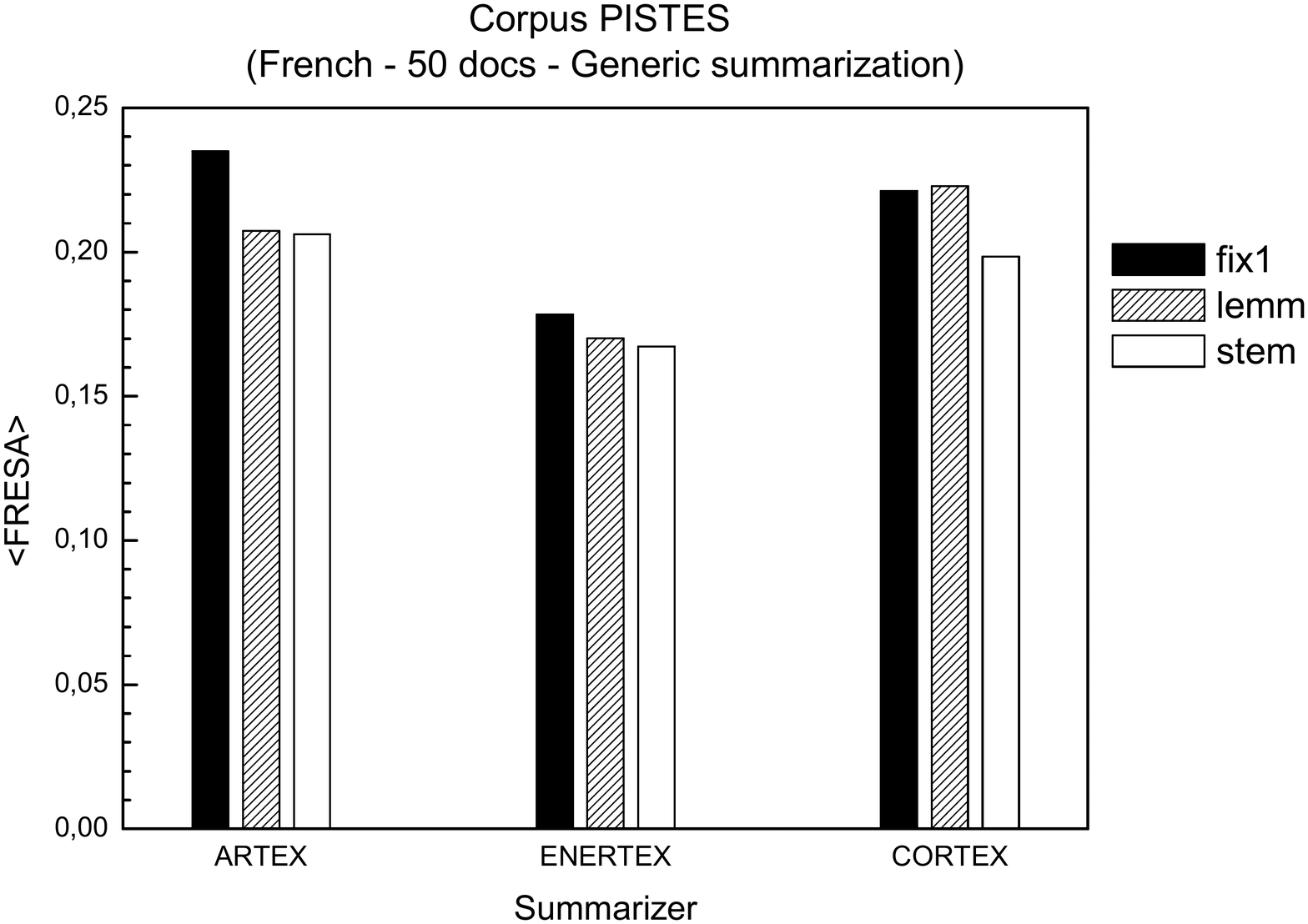}
	\caption{Histogram plot of content evaluation for French corpus {\sc Pistes} with $\langle$\textsc{Fresa}$\rangle$ scores for each summarizer.}
	\label{fig:pistes}
\end{figure}

\end{itemize}

\subsection{Processing Times Evaluation } 

Table \ref{tab:times} shows processing times for each corpus, 
following the normalization method for {\sc Cortex}, {\sc Artex} and {\sc Enertex} summarizers\footnote{All times are measured in a 7.8 GB of 
RAM computer, Core i7-2640M CPU @ 2.80GHz $\times$ 4 processor, running under 32 bits GNU/Linux (Ubuntu Version 12.04).}.
Processing times of ultra-stemming {\sc Fix$_1$} are shorter compared to all others methods.
By example, {\sc Cortex} is a very fast summarizer with $O(\log \rho^2)$ (where $\rho = P \times N$), and processing times for stemming and {\sc Fix$_1$} are close.
In other hand, {\sc Enertex} summarizer has a complexity of $O(\rho^2)$, then it needs more time to process the same corpus.
Performances of {\sc Artex} algorithm remain close to {\sc Cortex}.

\begin{table}[H]
\centering
\begin{tabular}{r|rrrr|}
	\hline
	& \multicolumn{4}{c}{\bf Summarizer Average Time}\\
	& \multicolumn{4}{c}{\bf (all corpora)}\\
	\hline
	\bf Normalization &\sc {\sc Cortex} & \textsc{\textsc{Artex}} & \sc Enertex \\
	\hline
		\textrm{Lemmatization} & 1.60' 	& 2.50' 	& 10.42'\\
		\textrm{Stemming}      & 0.54' 	& 1.29' 	&  9.47'\\
		\textsc{fix$_1$}   	  & \bf 0.32' & \sl 0.40' 	&  4.25' \\	
	\hline
\end{tabular}
\caption{Statistics of processing times (in minutes) of three summarizers over three corpora.}
\label{tab:times}
\end{table}

\section{Conclusions}
\label{sec:conclusion}

In this article we have introduced and tested a simple method for Automatic Text Summarization.
{\sc Artex} is a fast and very simple algorithm based in VSM model and the extractive paradigm.
The method uses a matrix representation to calculate a normalized score for each sentence, using the inner product of pseudo-(sentences|words) vectors.
The algorithm retains the salient information of each sentence of document.
An important aspect of our approach is that it does not requires linguistic knowledge or resources which makes it a simple and efficient summarizer method 
to tackle the issue of Automatic Text Summarization.

Summaries generated by {\sc Artex} system are pertinents.
The results obtained on corpora in English, Spanish and French show that {\sc Artex} can achieve good results for content quality.
Tests with other corpora (DUC and TAC evaluation campaigns, INEX, etc.) in mono-and multi-document guided by a subject, 
using content evaluation with (ROUGE evaluations) or without reference summaries still in progress.

\label{sect:bib}
\bibliographystyle{plain}
\bibliography{biblio}

\end{document}